\documentclass[aps,prd,onecolumn,groupedaddress,showpacs,nofootinbib,amssymb]{revtex4}
\usepackage[dvips]{graphicx}
\usepackage{amssymb}
\usepackage{amsmath}
\usepackage{graphicx,,color}
\usepackage{amsfonts}
\usepackage{bm}
\usepackage{cancel}
\usepackage{comment}
\usepackage[dvipsnames]{xcolor}
\usepackage{makecell}

\allowdisplaybreaks[4]

\begin{document}

\tolerance=5000

\title{$\mathcal{R}^2$-corrected Tachyon Scalar Field Inflation, the ACT Data, and Phantom Transition}
\author{S.D. Odintsov$^{1,2}$}\email{odintsov@ieec.cat}
\author{V.K. Oikonomou,$^{3,4}$}\email{voikonomou@gapps.auth.gr;v.k.oikonomou1979@gmail.com}
\affiliation{$^{1)}$ Institute of Space Sciences (ICE, CSIC) C. Can Magrans s/n, 08193 Barcelona, Spain \\
$^{2)}$ ICREA, Passeig Luis Companys, 23, 08010 Barcelona, Spain\\
$^{3)}$ Department of Physics, Aristotle University of
Thessaloniki, Thessaloniki 54124, Greece\\
$^{4)}$ Center for Theoretical Physics, Khazar University, 41
Mehseti Str., Baku, AZ-1096, Azerbaijan}

\begin{abstract}
Phantom divide line transitions are not possible in the context of
single scalar field scalar-tensor theories. In this article we
study a combined framework of a tachyonic minimally coupled single
scalar field theory in the presence of an $\mathcal{R}^2$
correction term and with a rescaled Einstein-Hilbert term of the
form $\sim \lambda \frac{\mathcal{R}}{16\pi G}$. Such terms can be
part of an $f(\mathcal{R})$ gravity which in the large curvature
regime yields such correction terms effectively. Alternatively,
such terms can simply be quantum corrections to the scalar field
action. We aim to answer two questions, firstly if this framework
can lead to phantom divide line transitions and secondly whether
the resulting model can be compatible with the ACT data. The model
we studied is an inverse square power-law model, well known from
tachyon inflation models. As we show, the field equations can be
cast in terms of the scalar field solely, however the resulting
theory is distinct from a single scalar field theory, because the
phantom divide line is crossed during inflation. Thus initially
the tachyonic nature of the scalar field generates a phantom
equation of state parameter, and during inflation the phantom
divide line is crossed, with the effective equation of state
parameter at the end of inflation being $w=-1/3$ which corresponds
to the non-accelerating state of the Universe. The model is proved
to be compatible with the ACT data, only when the gravity during
inflation is stronger than Einstein-Hilbert gravity, with the
effective gravitational constant during inflation being
$\frac{G}{\lambda}$. The effective theory is valid only during
inflation, thus Big-Bang nucleosynthesis is not affected by the
rescaling of the Einstein-Hilbert gravity. The feature of a
phantom crossing in $f(\mathcal{R},\phi)$ frameworks is new in the
literature.
\end{abstract}

\maketitle

\section{Introduction}

Inflation \cite{inflation1,inflation2,inflation3,inflation4} is
the mainstream description of the post-Planck era evolution of our
Universe. The inflationary description solves all the fundamental
issues of the hot Big Bang description, like the horizon problem,
the monopoles problem and the flatness problem and is a
theoretical necessity for a consistent evolution of the Universe,
not some artificially imposed theory that qualitatively describes
the physical laws of our Universe. Although inflation has not be
confirmed on solid grounds, the theory itself has indirectly been
confirmed by the facts that the spectrum of the primordial
perturbations is red tilted, the Universe is flat and the
perturbations are nearly Gaussian, facts that prove the quantum
origin of our Universe. Apart from that, the only consistent
theory that can explain successfully the large scale structure in
the Universe is inflation. Nevertheless, inflation must be
confirmed directly experimentally/observationally, and a smoking
gun observation that will confirm the occurrence of an
inflationary regime is the existence of a B-mode in the cosmic
microwave background (CMB) radiation. This will be the focus in
the CMB stage four experiments, like the Simons observatory
\cite{SimonsObservatory:2019qwx} and the LiteBird collaboration
\cite{LiteBIRD:2022cnt}. Apart from the CMB observations, a signal
that can also confirm the existence of an inflationary regime is a
stochastic gravitational wave background, and this will be probed
by future gravitational wave experiments
\cite{Hild:2010id,Baker:2019nia,Smith:2019wny,Crowder:2005nr,Smith:2016jqs,Seto:2001qf,Kawamura:2020pcg,Bull:2018lat,LISACosmologyWorkingGroup:2022jok}.
Such a stochastic gravitational wave background has already be
confirmed to exist at nanohertz frequencies from pulsar timing
array collaborations like NANOGrav \cite{NANOGrav:2023gor} but
inflation by itself cannot explain the signal
\cite{Vagnozzi:2023lwo,Oikonomou:2023qfz}.

An inflationary regime is realized in the context of general
relativity (GR) using a single scalar field
\cite{inflation1,inflation2,inflation3,inflation4}, however there
are other consistent realizations, like modified gravity
\cite{reviews1,reviews2,reviews3,reviews4}, which can harbor
consistently more exotic phenomena, that cannot be easily
described in the context of GR. One such example is the phantom
crossing recently pointed out by the DESI collaboration, but also
tensions and puzzles \cite{Cortes:2024lgw, Colgain:2024xqj,
Giare:2024smz,
Shlivko:2024llw,Han:2024qbw,You:2024hit,He:2024hmc,Tang:2024gtq,He:2020zns,Dai:2020rfo}.
Many descriptions have been proposed for this transition
\cite{OdintsovSGS:2024,Dai:2020rfo,He:2020zns,Nakai:2020oit,DiValentino:2020naf,Agrawal:2019dlm,Ye:2020btb,Vagnozzi:2021tjv,
Desmond:2019ygn,Hogas:2023pjz,OColgain:2018czj,Vagnozzi:2019ezj,
Krishnan:2020obg,Colgain:2019joh,Vagnozzi:2021gjh,Lee:2022cyh,Krishnan:2021dyb,Ye:2021iwa,Ye:2022afu,Verde:2019ivm,Menci:2024rbq,Adil:2023ara,
Reeves:2022aoi,Ferlito:2022mok,Vagnozzi:2021quy,DiValentino:2020evt,Sabogal:2024yha,DiValentino:2025sru,Odintsov:2025kyw,Odintsov:2025jfq,Kessler:2025kju,Nojiri:2025low}.
Such a phantom transition cannot occur in the context of GR at
all, unless $k$-essence theories are used \cite{kessencepaper},
plus a phantom era can be realized in the context of GR only if
phantom scalar fields are used. Motivated by this, in this work we
aim to study a rescaled phantom minimally coupled single scalar
field theory in the presence of $\mathcal{R}^2$ corrections. Such
theories emerge as effective $f(\mathcal{R})$ theories in the
large curvature regime, so during inflation. In addition to the
effective field theory approach that can explain the rescaling and
the presence of $\mathcal{R}^2$ corrections, the quantum
corrections of the minimally coupled single scalar field theory
contain, among other terms, the  $\mathcal{R}^2$ term and a
rescaled Einstein-Hilbert term \cite{Codello:2015mba}.
$\mathcal{R}^2$ corrected scalar field theories are studied in the
literature
\cite{Ema:2017rqn,Ema:2020evi,Ivanov:2021ily,Gottlober:1993hp,delaCruz-Dombriz:2016bjj,Enckell:2018uic,Kubo:2020fdd,Gorbunov:2018llf,Calmet:2016fsr,Oikonomou:2021msx,Oikonomou:2022bqb,Pi:2017gih,Salvio:2016vxi,Kannike:2015apa,Salvio:2015kka,Ozkan:2014cua},
but the approach is different, from what we shall adopt in this
article, since in the literature two scalar fields are used. In
this work we shall treat the resulting theory as a single scalar
field theory with quantum corrections expressed in terms of the
scalar field.

The aim in this work is two-fold, firstly we shall investigate
whether the tachyonic scalar field theory in the presence of
$\mathcal{R}^2$ corrections can achieve a phantom transition
primordially, and secondly, whether the resulting theory is
compatible with the ACT data. The ACT data
\cite{ACT:2025fju,ACT:2025tim} shook the ground of inflationary
physics, since the spectral index is constrained as follows,
\begin{equation}\label{act}
n_{\mathcal{S}}=0.9743 \pm
0.0034,\,\,\,\frac{\mathrm{d}n_{\mathcal{S}}}{\mathrm{d}\ln
k}=0.0062 \pm 0.0052\, .
\end{equation}
In addition, the updated Planck/BICEP tensor-to-scalar ratio
 is \cite{BICEP:2021xfz},
\begin{equation}\label{planck}
r<0.036\, ,
\end{equation}
at $95\%$ confidence. The ACT data initiated a large stream of
articles aiming to reconcile inflationary theories with the data
\cite{Kallosh:2025rni,Gao:2025onc,Liu:2025qca,Yogesh:2025wak,Yi:2025dms,Peng:2025bws,Yin:2025rrs,Byrnes:2025kit,
Wolf:2025ecy,Aoki:2025wld,Gao:2025viy,Zahoor:2025nuq,Ferreira:2025lrd,Mohammadi:2025gbu,Choudhury:2025vso,
Odintsov:2025wai,Q:2025ycf,Zhu:2025twm,Kouniatalis:2025orn,Hai:2025wvs,Dioguardi:2025vci,Yuennan:2025kde,
Kuralkar:2025zxr,Kuralkar:2025hoz,Modak:2025bjv,Oikonomou:2025xms,Odintsov:2025jky,
Aoki:2025ywt,Ahghari:2025hfy,McDonough:2025lzo,Chakraborty:2025wqn,NooriGashti:2025gug,Yuennan:2025mlg,
Deb:2025gtk,Afshar:2025ndm,Ellis:2025zrf,Iacconi:2025odq,Yuennan:2025tyx,Wang:2025cpp,Qiu:2025uot,Wang:2025dbj,Asaka:2015vza,Oikonomou:2025htz,Choudhury:2025hnu,Singh:2025uyr,Kim:2025dyi},
although caution is required interpreting the data
\cite{Ferreira:2025lrd}. Also in \cite{Peng:2026ofs} a program is
developed to produce ACT-compatible inflationary scenarios. The
theory we study in this work is a rescaled tachyonic minimally
coupled scalar field theory in the presence of $\mathcal{R}^2$
corrections, that uses the same spirit of Ref.
\cite{Oikonomou:2025htz}. By rescaled theory we mean that the
Einstein-Hilbert term primordially is not simply $\sim
\frac{\mathcal{R}}{16\pi G}$ but it contains a multiplication
parameter $\lambda$ and is of the form $\sim \lambda
\frac{\mathcal{R}}{16\pi G}$. Hence, gravity, as it is measured by
Newton's constant $G$, is stronger or weaker than Newton's
gravity, because the new gravitational constant is
$\frac{G}{\lambda}$. It turns out that the compatibility of the
model we shall consider occurs for a strong gravity, compared to
Newton's gravity. The model that achieves the compatibility with
the ACT data is an tachyonic inverse square power-law scalar field
model.

\section{Formalism of Rescaled Canonical Scalar Field Inflation with $R^2$ Corrections}

A scalar field action evaluated in its vacuum configuration,
\begin{equation}\label{generalscalarfieldaction}
\mathcal{S}_{\phi}=\int
\mathrm{d}^4x\sqrt{-g}\left(\frac{1}{2}Z(\phi)g^{\mu
\nu}\partial_{\mu}\phi
\partial_{\nu}\phi-\mathcal{V}(\phi)+h(\phi)\mathcal{R}
\right)\, .
\end{equation}
must have a minimal coupling to gravity or must be conformally
coupled. We consider the case that the scalar field is minimally
coupled to gravity, so $h(\phi)=1$ in Eq.
(\ref{generalscalarfieldaction}), but also we assume that the
scalar field is a tachyonic field, so $Z(\phi)=1$ and $h(\phi)=1$
in Eq. (\ref{generalscalarfieldaction}). The first quantum
corrections of the single scalar field action
(\ref{generalscalarfieldaction}) are \cite{Codello:2015mba},
\begin{align}\label{quantumaction}
&\mathcal{S}_{eff}=\int
\mathrm{d}^4x\sqrt{-g}\Big{(}\Lambda_1+\Lambda_2
\mathcal{R}+\Lambda_3\mathcal{R}^2+\Lambda_4 \mathcal{R}_{\mu
\nu}\mathcal{R}^{\mu \nu}+\Lambda_5 \mathcal{R}_{\mu \nu \alpha
\theta}\mathcal{R}^{\mu \nu \alpha \theta}+\Lambda_6 \square
\mathcal{R}\\ \notag &
+\Lambda_7\mathcal{R}\square\mathcal{R}+\Lambda_8 \mathcal{R}_{\mu
\nu}\square \mathcal{R}^{\mu
\nu}+\Lambda_9\mathcal{R}^3+\mathcal{O}(\partial^8)+...\Big{)}\, ,
\end{align}
where the parameters $\Lambda_i$, $i=1,2,...,6$ are dimensionful
constants. We only consider the terms $\mathcal{R}$ and
$\mathcal{R}^2$ in this article and we shall analyze the physics
that these two induce on the tachyonic inflation. Thus the full
gravitational action which we consider in this work is the
following,
\begin{equation}
\label{action} \centering
\mathcal{S}=\int{d^4x\sqrt{-g}\left(\frac{ \lambda
\mathcal{R}+\frac{\mathcal{R}^2}{36M^2}}{2\kappa^2}+\frac{1}{2}g^{\mu
\nu}\partial_{\mu}\phi
\partial_{\nu}\phi-\mathcal{V}(\phi)\right)}\, ,
\end{equation}
with $\kappa^2=8\pi G=\frac{1}{M_p^2}$ and $M_p$ is the reduced
Planck mass. Also $M$ is a mass scale the value of which will be
determined by the correct inflationary phenomenology. The action
(\ref{action}) describes a rescaled version of the
Einstein-Hilbert gravity since the term $\sim \mathcal{R}$
involves a non-trivial multiplication factor $\lambda$. The theory
thus has a different gravitational constant equal to
$\frac{G}{\lambda}$ due to the fact that the rescaled
Einstein-Hilbert term has the form $\sim \lambda
\frac{\mathcal{R}}{16\pi G}$. Hence, gravity can be weaker or
stronger than the Einstein-Hilbert gravity, depending on the
values of the rescaling parameter $\lambda$. Note that the term
$\lambda \mathcal{R}$ is an effective term active only during the
inflationary era, and for large curvatures. This can be the result
of a total $f(\mathcal{R})$ gravity of the form
\cite{Oikonomou:2025qub},
\begin{equation}\label{frini}
f(\mathcal{R})=\mathcal{R}+\frac{\mathcal{R}^2}{36M^2}-\gamma
\theta  \Lambda -\theta \mathcal{R} \exp \left(-\frac{\gamma
\Lambda }{\mathcal{R}}\right)-\frac{\Lambda
\left(\frac{\mathcal{R}}{m_s^2}\right)^{\delta }}{\zeta }\, ,
\end{equation}
with $\Lambda$ being the cosmological constant, and
$m_s^2=\frac{\kappa^2 \rho_m^{(0)}}{3}$, with $\rho_m^{(0)}$
denoting the energy density of cold dark matter at the present
time epoch. The exponential term of the $f(\mathcal{R})$ gravity
of Eq. (\ref{frini}), in the large curvature regime becomes,
\begin{equation}\label{expapprox}
\theta  \mathcal{R} \exp \left(-\frac{\gamma  \Lambda
}{\mathcal{R}}\right)\simeq-\gamma \theta  \Lambda -\frac{\gamma
^3 \theta \Lambda^3}{6 \mathcal{R}^2}+\frac{\gamma ^2 \theta
\Lambda ^2}{2 \mathcal{R}}+\theta \mathcal{R}\, ,
\end{equation}
hence, the complete effective action in the large curvature regime
becomes,
\begin{equation}\label{effectiveaction}
\mathcal{S}=\int
d^4x\sqrt{-g}\left(\frac{1}{2\kappa^2}\left(\lambda
\mathcal{R}+\frac{\mathcal{R}^2}{36M^2}+ \frac{\gamma ^3 \theta
\Lambda ^3}{6 \mathcal{R}^2}-\frac{\gamma ^2 \theta \Lambda ^2}{2
\mathcal{R}}-\frac{\Lambda}{\zeta
}\left(\frac{\mathcal{R}}{m_s^2}\right)^{\delta
}+\mathcal{O}(1/\mathcal{R}^3)+...\right)+\frac{1}{2}g^{\mu\nu}\partial_\mu\phi\partial_\nu\phi-\mathcal{V}(\phi)\right)\,
,
\end{equation}
with $\lambda=1-\theta$. Note that post-inflationary, the large
curvature expansion is not valid anymore, and thus one returns to
the $f(\mathcal{R})$ of Eq. (\ref{frini}), hence the Big Bang
Nucleosynthesis physics is not affected, Newton's constant is $G$
post-inflationary. For a flat Friedmann-Robertson-Walker (FRW)
spacetime,
\begin{equation}
    \centering\label{metric}
    \mathrm{d}s^2 = - \mathrm{d} t^2 + a(t) \sum_{i = 1}^3 \mathrm{d} x_i^2,
\end{equation}
the field equations of the tachyonic rescaled
$\mathcal{R}^2$-corrected scalar field gravitational action of Eq.
(\ref{action}) are,
\begin{equation} \label{Friedmann}
    3 f_{\mathcal{R}} \mathcal{H}^2=\frac{R f_{\mathcal{R}} -f}{2}-3\mathcal{H} \dot F_{\mathcal{R}}+\kappa^2\big(-\frac{1}{2}\dot\phi^2+\mathcal{V}(\phi)\big) \ ,
\end{equation}
\begin{equation} \label{Raychad}
    -2 f_{\mathcal{R}} \dot{\mathcal{H}} = -\kappa^2 \dot\phi^2 + \ddot f_{\mathcal{R}} -\mathcal{H} \dot f_{\mathcal{R}} \ ,
\end{equation}
\begin{equation} \label{fieldeqmotion}
   \ddot\phi+3\mathcal{H}\dot\phi-\mathcal{V}'=0 \ ,
\end{equation}
where the ``dot'' denotes derivatives with respect to the cosmic
time, while the ``prime'' is denoting differentiation with respect
to the scalar field $\phi$ and finally,
$f_{\mathcal{R}}=\frac{\partial f}{\partial \mathcal{R}}$. By
taking into account that $f(\mathcal{R})=\lambda
\mathcal{R}+\frac{\mathcal{R}^2}{36M^2}$ and also that for the
flat FRW metric, the Ricci scalar and its first derivative have
the following forms,
\begin{equation}\label{ricciscalar}
    \mathcal{R} =12 \mathcal{H}^2 + 6\dot{\mathcal{H}} \, , \ \dot{\mathcal{R}}=24\mathcal{H}\dot{\mathcal{H}} +6\ddot{\mathcal{H}},
\end{equation}
the field equations acquire the following form,
\begin{equation} \label{Friedman3}
    3\lambda\mathcal{H}^2+\frac{\mathcal{H}^2}{M^2}\dot{\mathcal{H}}=\kappa^2\mathcal{V}(\phi) ,
\end{equation}
\begin{equation} \label{Raycha3}
    -2 \lambda \dot{\mathcal{H}}-\frac{2}{M^2}\dot{\mathcal{H}}^2 =-\kappa^2 \dot \phi^2,
\end{equation}
\begin{equation} \label{dotphi}
   \dot \phi \simeq \frac{\mathcal{V}'}{3\mathcal{H}}\, .
\end{equation}
Note that we assumed a slow-roll era,
\begin{equation}\label{slowrollH}
    \dot{\mathcal{H}} \ll \mathcal{H}^2 \ , \ \ddot{\mathcal{H}} \ll \mathcal{H} \dot{\mathcal{H}},
\end{equation}
and in addition that the following conditions hold true,
\begin{equation}\label{approxH}
    \frac{\dot{\mathcal{H}}^2}{M^2} \ll \mathcal{H}^2,\,\,\,\frac{\dot{\mathcal{H}}^2}{M^2} \ll
    \mathcal{V}(\phi)\, .
\end{equation}
The validity of the above approximations must be examined at the
first horizon crossing, when the phenomenology of a specific model
is examined. The Raychaudhuri equation is basically a second order
polynomial equation with respect to $\dot{\mathcal{H}}$, with
solution,
\begin{equation}\label{dothanalyticsol}
\dot{\mathcal{H}}=\frac{-M^2\lambda+M\sqrt{M^2\lambda^2+2\dot{\phi}^2\kappa^2}}{2}\,
.
\end{equation}
We shall furthermore make the assumption,
\begin{equation}\label{taylorphi}
    \frac{2\kappa^2 \dot \phi^2}{M^2} \ll 1\, ,
\end{equation}
which will also be tested when the phenomenology of a model is
examined. In view of Eq. (\ref{taylorphi}), the field equations at
leading order can be written,
\begin{equation} \label{Friedman}
    \lambda \mathcal{H}^2 \simeq \frac{\kappa^2 \mathcal{V}(\phi)}{3}+\mathcal{O}(\frac{\kappa^2 \dot{\phi}^2}{2M^2}\mathcal{H}^2),
\end{equation}
\begin{equation} \label{Raycha}
     \dot{\mathcal{H}} \simeq \frac{\kappa^2 \dot{\phi}^2}{2\lambda}  -\frac{\kappa^4
     \dot{\phi}^4}{4\lambda^3 M^2}\, ,
\end{equation}
and we used,
\begin{equation}\label{expansionref}
\sqrt{\lambda ^2+x}\simeq \lambda -\frac{x^2}{8 \lambda
^3}+\frac{x}{2 \lambda }\, .
\end{equation}
The equations (\ref{Friedman}), (\ref{Raycha}) and (\ref{dotphi})
quantify the quantum-corrected minimally coupled tachyonic scalar
field action during inflation, or equivalently the
$f(\mathcal{R})$ gravity effective minimally coupled tachyonic
scalar field action during inflation. This is effectively an
$f(\mathcal{R},\phi)$ theory of gravity, so for these theory the
slow-roll indices are \cite{Hwang:2005hb},
\begin{align}\label{slowrollindices}
\epsilon_1=-\frac{\dot{\mathcal{H}}}{\mathcal{H}^2},\,\,\,\epsilon_2=\frac{\ddot
\phi }{\mathcal{H} \dot \phi},\,\,\,
\epsilon_3=\frac{\dot f_{\mathcal{R}} }{2\mathcal{H}f_{\mathcal{R}}},\,\,\,\epsilon_4=\frac{\dot E}{2\mathcal{H}E}\, ,\\
\end{align}
with,
\begin{equation}\label{E1}
    E=f_{\mathcal{R}}+\frac{3 \dot f_{\mathcal{R}}^2}{3\kappa^2\dot
    \phi^2}\, .
\end{equation}
Combining Eqs. (\ref{Friedman}) and (\ref{Raycha}), the first
slow-roll index $\epsilon_1$ becomes,
\begin{equation}\label{e1}
\epsilon_1=-\frac{\lambda}{2\kappa^2}\left(\frac{\mathcal{V}'}{\mathcal{V}}\right)^2
+\frac{1}{12M^2}{\left(\frac{\mathcal{V}'}{\mathcal{V}}\right)}^2\frac{\mathcal{V}'^2}{\mathcal{V}}\,
.
\end{equation}
and in addition, the second slow-roll index is,
\begin{equation}\label{e2}
\epsilon_2=\frac{\mathcal{V}''}{\kappa^2 \mathcal{V}}+\epsilon_1
\, .
\end{equation}
Furthermore, the third slow-roll index $\epsilon_3$ takes the
form,
\begin{equation}\label{e3}
\epsilon_3=\frac{\epsilon_1}{-1-\frac{3\lambda
M^2}{2\mathcal{H}^2}+\frac{\epsilon_1}{2}} \, ,
\end{equation}
and in addition, $E$ and $\dot E$ are,
\begin{equation}\label{E}
E=-\frac{\lambda}{\kappa^2}+\frac{4\mathcal{H}^2}{3M^2\kappa^2}+\frac{3}{2\kappa^4\dot{\phi}^2}\left(\frac{4\mathcal{H}\dot{\mathcal{H}}}{3M^2}
\right)^2\, ,
\end{equation}
\begin{equation}\label{dotE}
\dot E=\frac{4 \mathcal{H} \dot{\mathcal{H}}}{3M^2} +
\frac{16}{\kappa^4\dot{\phi}^4M^4}\left(\mathcal{H}\dot{\mathcal{H}}^3\dot{\phi}^2-\mathcal{H}^2\dot{\mathcal{H}}^2\dot{\phi}\ddot{\phi}
\right),
\end{equation}
hence the slow-roll index $\epsilon_4$ can be obtained. The
observational indices, and specifically, the spectral index of the
scalar curvature perturbations is \cite{Hwang:2005hb},
\begin{equation}\label{ns}
n_\mathcal{S}= 1 - \frac{4\epsilon_1 + 2\epsilon_2 -2 \epsilon_3 +
2 \epsilon_4}{
 1 -\epsilon_1} \, ,
\end{equation}
while the tensor-to-scalar ratio, is
\begin{equation}\label{r}
r=16|\epsilon_1 + \epsilon_3|\, .
\end{equation}
In addition, the $e$-foldings number for the rescaled scalar
theory is,
\begin{equation}\label{N}
N=\frac{\kappa^2}{\lambda}\int_{\phi_f} ^{\phi_i}
\frac{\mathcal{V}}{\mathcal{V}'} d \phi .
\end{equation}
Also, another complication arises for the combined
$f(\mathcal{R},\phi)$ theory at hand, the amplitude of scalar
perturbations $\mathcal{P}_{\zeta}(k_*)$, is not trivially
affected by the multiplication constant of the scalar potential,
but on the contrary it is much more complicated. Specifically, its
definition in terms of the slow-roll indices is
\cite{Hwang:2005hb},
\begin{equation}\label{powerspectrumscalaramplitude}
\mathcal{P}_{\zeta}(k)=\left(\frac{k \left((-2
\epsilon_1-\epsilon_2-\epsilon_4) \left(0.57\, +\log \left(\left|
\frac{1}{1-\epsilon_1}\right| \right)-2+\log
(2)\right)-\epsilon_1+1\right)}{2 \pi  z}\right)^2\, ,
\end{equation}
where $z=\frac{(\dot{\phi} k) \sqrt{\frac{E(\phi
)}{f_{\mathcal{R}/\kappa^2 }}}}{\mathcal{H}^2 (\epsilon_3+1)}$,
and in addition for the tachyonic $f(\mathcal{R},\phi)$ theory at
hand, we have $k=a\mathcal{H}$ evaluated at the first horizon
crossing where it also holds true that
$\eta=-\frac{1}{a\mathcal{H}}\frac{1}{-\epsilon_1+1}$. Note again
that the sound wave speed is trivial, since we are essentially
considering a tachyonic $f(\mathcal{R},\phi)$. There are
constraints on the amplitude of the scalar perturbations from the
Planck data \cite{Planck:2018jri} which indicate that
$\mathcal{P}_{\zeta}(k_*)=2.196^{+0.051}_{-0.06}\times 10^{-9}$.
With the formalism of the rescaled effective
$\mathcal{R}^2$-corrected tachyonic scalar field theory at hand,
in the next section, we shall examine the phenomenology of a
specific model which will turn out to be ACT-compatible.

\section{Rescaled $\mathcal{R}^2$-corrected Inverse Square Power-law Inflation: Phenomenology and Phantom Crossing}

We apply the formalism developed in the previous section for the
following inverse square power-law scalar potential,
\begin{equation}\label{V1}
    \mathcal{V}(\phi)=\frac{\mathcal{V}_0}{\kappa^4}\frac{1}{\left(\kappa \phi\right)^2},
\end{equation}
where $\mathcal{V}_0$ being a dimensionless parameter. It will
prove that the above model in the context of the rescaled
effective $\mathcal{R}^2$-corrected tachyonic scalar field theory
will be ACT-compatible. Remarkably, the inverse square power-law
models are used in tachyonic theories of inflation
\cite{Feinstein:2002aj,Sami:2002zy}, but these potentials are also
found in ordinary inflationary frameworks \cite{Lu:2013roa}. For
this scalar potential, and by using Eqs. (\ref{e1}), (\ref{e2}),
(\ref{e3}),(\ref{ns}), (\ref{r}), (\ref{N}), the first slow-roll
index $\epsilon_1$ takes the form,
\begin{equation}\label{e1V1}
    \epsilon_1=\frac{4 \mathcal{V}_0}{3 \kappa ^8 M^2 \phi ^6}-\frac{2 \lambda }{\kappa ^2 \phi ^2}\, ,
\end{equation}
thus by solving $\epsilon_1\simeq \mathcal{O}(1)$ we get the value
of the scalar field when inflation ends,
\begin{equation}\label{scalarfieldendofinflation}
\phi_f=\frac{\sqrt{-\frac{2 \lambda }{\kappa ^2}+\frac{\sqrt[3]{-8
\kappa ^{18} \lambda ^3 M^6+18 \kappa ^{16} M^4 \mathcal{V}_0+6
\sqrt{S(\phi )}}}{\kappa ^8 M^2}+\frac{2\ 2^{2/3} \kappa ^4
\lambda ^2 M^2}{\sqrt[3]{-4 \kappa ^{18} \lambda ^3 M^6+9 \kappa
^{16} M^4 \mathcal{V}_0+3 \sqrt{S(\phi )}}}}}{\sqrt{3}}\, ,
\end{equation}
where $S(\phi)$ is defined as follows,
\begin{equation}\label{sphiaux}
S(\phi)=\kappa ^{32} \left(-M^8\right) \mathcal{V}_0 \left(8
\kappa ^2 \lambda ^3 M^2-9 \mathcal{V}_0\right)
\end{equation}
\begin{figure}
\centering
\includegraphics[width=25pc]{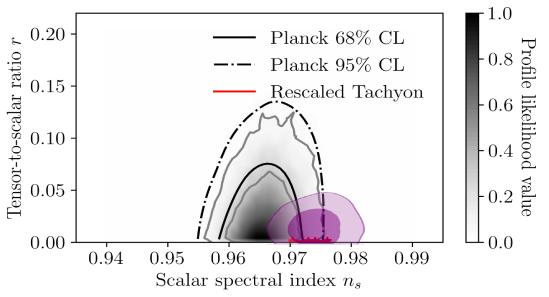}
\caption{The 2018 marginalized Planck likelihood curves, the ACT
constraints and the updated Planck constraints on the
tensor-to-scalar ratio, versus the rescaled
$\mathcal{R}^2$-corrected tachyonic inverse square power-law
inflation model for $\mathcal{V}_0=8\times 10^{-12}$,
$\beta=0.000004$, $\lambda=0.1$, and $N$ in the range
$N=[50,60]$.}\label{plot1}
\end{figure}
In the same vain, we the second slow-roll index reads,
\begin{equation}\label{epsilon2new}
\epsilon_2=\frac{6-2 \lambda }{\kappa ^2 \phi ^2}+\frac{4
\mathcal{V}_0}{3 \kappa ^8 M^2 \phi ^6}
\end{equation}
and also $\epsilon_3$ reads,
\begin{equation}\label{epsilon3new}
\epsilon_3=\frac{4 \mathcal{V}_0 \left(2 \mathcal{V}_0-3 \kappa ^6
\lambda M^2 \phi ^4\right)}{-27 \kappa ^{12} \lambda ^2 M^4 \phi
^8-6 \kappa ^6 M^2 \mathcal{V}_0 \phi ^4 \left(\kappa ^2 \phi
^2+\lambda \right)+4 \mathcal{V}_0^2}\, .
\end{equation}
The slow-roll index  $\epsilon_4$ can also be evaluated, but it is
too complicated to quote it here. Now, by using Eq. (\ref{N}) and
also having in mind the value of the scalar field when inflation
ends, that is $\phi_f$, using $N=-\frac{\kappa ^2 \phi_f ^2}{4
\lambda }+\frac{\kappa ^2 \phi_i ^2}{4 \lambda }$, we get
$\phi_i$,
\begin{equation}\label{phianalytic}
\phi_i=\frac{2 \sqrt{2} \sqrt{\Gamma  \lambda +\lambda  N}}{\kappa
}\, ,
\end{equation}
where $\Gamma$ is defined as follows,
\begin{equation}\label{gammadefinition}
\Gamma=\frac{\kappa ^2 \phi_f^2}{4 \lambda }\, .
\end{equation}
Accordingly, the scalar spectral index of the primordial
perturbations can be evaluated and the tensor-to-scalar ratio, but
their expressions are too lengthy to quote these here. So let us
proceed with the phenomenology of the inverse square power-law
model (\ref{V1}). By substituting for simplicity $M=\beta/\kappa$
and by taking $\mathcal{V}_0=4.4\times 10^{-9}$, $\beta=6\times
10^{-9}$ and $\lambda=0.5$, for $N=60$ $e$-foldings we get
$n_{\mathcal{S}}=0.97682$, $r=0.0012151$ and in addition
$\mathcal{P}_{\zeta}(k)= 2.15379 \times 10^{-9}$. Therefore, the
inverse square model is compatible with both the ACT data and the
updated BICEP/Planck data. In Fig. \ref{plot1} we present the
confrontation of the rescaled $\mathcal{R}^2$-corrected tachyonic
inverse square power-law inflation model of Eq. (\ref{V1}) with
the ACT, the Planck 2018 and the updated Planck/BICEP data for
$\mathcal{V}_0=4.4\times 10^{-9}$, $\beta=6\times 10^{-9}$ and
$\lambda=0.5$ and $N$ in the range $N=[50,60]$, taking also into
account the constraints on the amplitude of the scalar
perturbations. As it is obvious, the model is well fitted in the
ACT and  Planck/BICEP constraints, thus the model is overall
viable. Also note that the viability of the model comes for
$\lambda=0.5$ so basically a stronger gravity than
Einstein-Hilbert gravity makes the model compatible with the ACT
data. Now let us check explicitly whether the approximations we
made earlier when deriving the field equations hold true. Indeed,
one can check that the approximation of Eq. (\ref{taylorphi})
holds true since at first horizon crossing, that is for
$\phi=\phi_i$, and $\mathcal{V}_0=4.4\times 10^{-9}$,
$\beta=6\times 10^{-9}$, $\lambda=0.5$, and $N=60$ we have,
$\frac{2\kappa^2 \dot
\phi^2}{M^2}/H^2\Big{|}_{\phi=\phi_i}=0.000217087$. It can also be
checked that the slow-roll indices are smaller than unity during
the inflationary regime, and in addition,
$\frac{\kappa^2\dot{\phi}^2}{M^2}\Big{|}_{\phi=\phi_i}=0.096854$.
Hence all the approximations we made are valid. Before closing,
let us discuss the important issue of the behavior of the
effective equation of state (EoS) parameter during inflation. The
question is whether initially, at first horizon crossing, the EoS
parameter is phantom of quintessential. The reason for this
question is basically the existence of the $\mathcal{R}^2$ term
and how it affects the EoS parameter. So basically, does the
tachyonic nature of the scalar field affects the EoS parameter or
does the $\mathcal{R}^2$ term overwhelms the EoS parameter? The
EoS parameter for the $f(\mathcal{R},\phi)$ theory at hand is,
\begin{equation}\label{EoS}
w_{eff}=-1-\frac{2}{3}\frac{\dot{\mathcal{H}}}{\mathcal{H}^2}\, ,
\end{equation}
and we will be using the expressions for $\dot{\mathcal{H}}$ and
$\mathcal{H}$ quoted in Eqs. (\ref{Friedman}) and (\ref{Raycha}).
We shall evaluate the EoS at the beginning of the inflationary
era, thus at $\phi=\phi_i$, and also at the end of the
inflationary era, so at $\phi=\phi_f$. Thinking normally, one
expects that the inflationary era will start from a tachyonic EoS
due to the tachyon scalar field, and will end at a
non-acceleration state with $w_{eff}=-1/3$ at the end of
inflation. This is not certain though due to the presence of the
$\mathcal{R}^2$ term, which may provide a quintessential quasi-de
Sitter EoS. Thus we must evaluate the EoS numerically using Eqs.
(\ref{EoS}), (\ref{Friedman}) and (\ref{Raycha}) and evaluate the
EoS at the beginning of inflation, when $\phi=\phi_i$ and at the
end of inflation, at $\phi=\phi_f$. So for the model at hand, by
using $\mathcal{V}_0=4.4\times 10^{-9}$, $\beta=6\times 10^{-9}$
and $\lambda=0.5$, and $N=60$ we get $w_{eff}=-1.0018$ at the
beginning of inflation at $\phi=\phi_i$ and $w_{eff}=-0.3333$ at
the end of inflation, that is at $\phi=\phi_f$. Thus, remarkably,
a phantom divide line is crossed primordially from the rescaled
$\mathcal{R}^2$-corrected tachyonic inverse square power-law
inflation model. This transition would be impossible to achieve in
the context of a pure scalar field theory \cite{kessencepaper},
but it is possible in the context of rescaled
$\mathcal{R}^2$-corrected tachyonic inverse square power-law
inflation model, and this is possibly due the presence of the
$\mathcal{R}^2$ term.

We need to note that the inverse phantom transition we described
in this model occurs only during the inflationary era. So we
highlight the possibility of having phantom scalar fields in the
presence of effective $f(\mathcal{R})$ gravity terms, which
realizes the inverse phantom transition. This result is
reminiscent of the late-time transition pointed out by the DESI
data, however our result does not necessarily imply that something
similar could happen at late times. One is required to check this
possibility explicitly for a specific model of phantom scalars and
effective $f(\mathcal{R})$ gravity terms.

\section{Conclusions}

In this article we studied the inflationary phenomenology of a
rescaled tachyonic minimally coupled scalar field theory with
$\mathcal{R}^2$ corrections. Our aim was two-fold, firstly to see
whether a phantom divide line transition can occur in this
tachyonic scalar theory and secondly whether the resulting model
can be compatible with the ACT data. As we demonstrated, this is
possible in the theoretical framework of tachyonic minimally
coupled scalar field theory with $\mathcal{R}^2$ corrections.
Specifically, the scalar model that achieved both aims is an
inverse square power-law model, which is used in tachyon inflation
physics \cite{Feinstein:2002aj,Sami:2002zy}. We found that the
viability with the ACT data occurs only for a stronger gravity
than Einstein-Hilbert gravity. The rescaling of the
Einstein-Hilbert term in addition to the $\mathcal{R}^2$
corrections can occur in single scalar field theory if the theory
contains $f(\mathcal{R})$ gravity which effectively in the large
curvature regime can yield both the rescaling of the
Einstein-Hilbert term and the $\mathcal{R}^2$ correction term, see
for example \cite{Oikonomou:2025qub}, for more general theories of
this sort. Regarding the rescaling, it is of the form $\sim
\lambda \frac{\mathcal{R}}{16\pi G}$, thus the gravitational
constant is of the form $\frac{G}{\lambda}$, hence gravity can be
weaker or stronger than GR, depending on the value of the
rescaling parameter $\lambda$. The model we studied in this
article becomes ACT-compatible for stronger gravity than
Einstein-Hilbert gravity. Our investigation also indicated that
the model is severely parametrically restricted by the amplitude
of the scalar perturbations, since the resulting theory is not a
simple scalar-tensor theory, but an effective quantum corrected
theory, hence the amplitude of the scalar perturbations is not
only affected by the multiplication constant parameter of the
scalar potential, but it is much more complicated. Furthermore,
two important questions in the theoretical framework we studied
are whether the EoS of the $f(\mathcal{R},\phi)$ system initially
was a phantom one, and secondly if the phantom divide line is
crossed during inflation. These are natural questions to ask,
since initially the cosmological fluid contained a tachyon and an
$\mathcal{R}^2$ correction term, so it is not certain whether the
tachyon or the $\mathcal{R}^2$ correction term initially dominated
the EoS, and of course, since inflation ends eventually, this
means that the EoS parameter will reach the value $w=-1/3$. As we
showed, initially the EoS was phantom, and also that the
cosmological dynamics lead to a phantom divide line crossing, a
feature which is impossible in single field scalar-tensor
frameworks. Thus, the presence of the $\mathcal{R}^2$ eventually
made the phantom divide crossing possible. Hence, although the
field equations are written in terms of a scalar field, the
dynamics of the combined tachyonic scalar theory with
$\mathcal{R}^2$ corrections are quite distinct from a
scalar-tensor theory. We believe that this is a new feature in the
context of $f(\mathcal{R},\phi)$ theory that has never appeared in
the literature. Note that phantom theories where phantom crossing
did not occur, may eventually lead to future singularities, for a
review on finite-time singularities, see \cite{deHaro:2023lbq}.

\end{document}